\documentclass[preprint,12pt,prd,amsmath,amssymb,floatfix]{revtex4}

\usepackage{bbold}
\usepackage{bm}
\usepackage{color}
\usepackage{dcolumn}
\usepackage{feynmf} 
\usepackage{graphics}
\usepackage{graphicx}
\usepackage{subfigure}
\usepackage{slashed}
\usepackage{placeins}

\usepackage[colorlinks=true,urlcolor=black]{hyperref}

\hypersetup{backref = true,pagebackref = true,hyperindex = true, colorlinks = true,
  breaklinks = true, urlcolor = blue, linkcolor = blue, bookmarks = true,
  bookmarksopen = true, citecolor=red}

\def\al{\alpha}

\def\veps{\varepsilon}

\def\be{\begin{equation}}
\def\ee{\end{equation}}
\def\bea{\begin{eqnarray}}
\def\eea{\end{eqnarray}}
\def\bse{\begin{subequations}}
\def\ese{\end{subequations}}
\def\bc{\begin{center}}
\def\ec{\end{center}}

\def\ra{\rightarrow}

\def\nonum{\nonumber}

\newcommand{\comment}[1]{}

\catcode`,\active

\catcode`\,12

\begin{document}

\title{Short review of  interaction effects in graphene}

\author{ A.\ V.~Kotikov
}
\affiliation{
  Bogoliubov Laboratory of Theoretical Physics, Joint Institute for Nuclear Research, 141980 Dubna, Russia.\\
}

\date{\today}

\begin{abstract}

  We review field theoretical studies dedicated to understanding the effects of electron-electron interaction in graphene, which is characterized by gapless bands, strong electron-electron interactions, and emerging Lorentz invariance deep in the infrared.
  We consider the influence of interactions on the transport properties of the system as well as their supposedly decisive influence on the potential dynamical generation of a
gap.

\end{abstract}

\maketitle

\begin{fmffile}{fmfmqft-18}

Low-energy excitations of graphene have a linear relativistic spectrum.
Thus, a low-energy efficient description of graphene may well be done by analogy with Quantum Electrodynamics (QED),
as Semenoff \cite{Semenoff:1984dq} foresaw a long time ago.
However, there are important differences from conventional relativistic QED. First, electrons in graphene propagate at the Fermi velocity
$v$,
which is much less than the speed of light: $v = c/300$. In the presence of interactions, this means that the Lorentz invariance is lost.
The second difference with respect to
conventional QED is that we are dealing with fermions in the spatial dimensions $D_e=2$ interacting through the
gauge field in the spatial dimensions $D_\gamma = 3$.
Thus, this system can be seen as a physical realization of a ``brane-like'' universe, similar to those often used in particle physics for a
larger (and non-physical) number of dimensions of space.
Thirdly, we are dealing with a theory at
strong coupling, since
\be
\alpha_g = \frac{e^2}{4 \pi \hbar \kappa v} = \frac{2.2}{\kappa}\, ,
\label{intro:alg}
\ee
where the dielectric constant $\kappa \approx 1$ in the case of suspended graphene.

An important part of our research is the appearance of Lorentz invariance at low energies in graphene, which was theoretically proved
in [\onlinecite{Gonzalez:1993uz}] using the RG study: $v \ra v^*=c $ and
$\al_g \ra \al_g^* = \al=1/137$.
Although {\it a priori} this limit is primarily of academic interest, the general motivation for considering reduced QED models (with $D_e=2$ and
$D_\gamma = 3$)
is that relativistic invariance allows a rigorous and systematic study of interaction effects.
Indeed, the whole powerful multiloop approach (see, e.g., [\onlinecite{Kotikov:2018wxe}]) combined with the field theory renormalization method, originally developed in particle physics and statistical mechanics, can be applied in this case.

{\bf 1.} Consider first the optical conductivity. It turns out that, despite the strength and range of the Coulomb interaction between charge carriers, the available experimental results are consistent with the case of free fermions, with possible deviations of the order of $2\%$. Experiments
did not even unambiguously associate these deviations with interactions (see Ref.
[\onlinecite{Gusynin:2006ym}] and discussions therein).
Extensive theoretical attempts have been made to understand such small corrections to the
optical conductivity due to interactions (see e.g.,
[\onlinecite{Teber:2014ita,Teber:2018jdh}] and references therein). Analytically, this problem is usually considered using perturbation theory:
\be
\sigma = \sigma_0\,\big( 1 + \mathcal{C}_1 \al_g
+ \cdots \big)
= \sigma_0\,\big( 1 + \mathcal{C}_1^* \al + \cdots \big),
\label{intro:sigma}
\ee
where $\sigma_0 = e^2/(4\hbar)$ is the free fermion conductivity~\cite{Fradkin86.PhysRevB.33.3263} and $\mathcal{C}_1$ and
$\mathcal{C}_1^*$ are interaction correction coefficients, evaluated in the cases $v=0$ and $v=c$, respectively. They are in good {\it quantitative} agreement with each other:
\be
\mathcal{C}_1 = \frac{19-6\pi}{12} \approx 0.013, \qquad \mathcal{C}_1^* = \frac{92-9\pi^2}{18\pi} \approx 0.056\, .
\label{models:C-res}
\ee
It is necessary to comment on the coefficient $\mathcal{C}_1$ (see also Refs. [\onlinecite{Teber:2014ita,Teber:2018jdh}]).
Exact calculations of the corresponding two-loop diagrams in dimensional regularization give the value:

\be
\mathcal{C}_1^{(\text{bare})} = \frac{11-3 \pi}{6}\, .
\nonum
\ee
Then, at the level of renormalization, it is necessary to subtract the contribution of the divergent subgraph associated with the renormalization of the Fermi velocity.
It is convenient to do this using the Bogolyubov-Parasyuk-Hepp-Zimmerman (BPHZ) procedure
\cite{Bogoliubov:1957gp} for renormalization,
\footnote{Usually the BPHZ procedure leads to the local form of singularities of Feynman diagrams, i.e. to their independence from external momenta (see, e.g.,
  [\onlinecite{Kotikov:2021hsy}]). The case of the optical conductivity is rather unusual, since the singularity of the subgraph (i.e. $1/\veps$ in dimensional
  regularization)
  is compensated by the contribution $\sim \veps$ of the second integration. So the two-loop result for $\mathcal{C}_1^{(\text{bare})}$ is finite ($=(11-3 \pi)/6$). But since the subgraph
  is singular, $\mathcal{C}_1^{(\text{bare})}$ should be supplemented with a counter-term related to subgraph renormalization. The corresponding pole is also compensated by the
  contribution $\sim \veps$ from the integration of the remainder. Thus, this additional contribution is also finite ($=1/4$) and must be subtracted from the results of
  calculating the two-loop diagrams as shown in Eq. (\ref{C1}).}
which gives
~\cite{Teber:2018jdh}
\be
\mathcal{C}_1 = \frac{11-3 \pi}{6} - \frac{1}{4} = \frac{19-6\pi}{12} \approx 0.013\, ,
\label{C1}
\ee
according to earlier studies \cite{Mishchenko2008} Mishchenko. The calculations \cite{Teber:2014ita,Teber:2018jdh} demonstrate the efficiency and power of multiloop methods using dimensional
regularization in the MS-scheme along with the use of the BPHZ renormalization recipe for nonrelativistic calculations.

{\bf 2.} The second issue we will focus on concerns the dynamic gap generation in graphene.
For relativistic systems, this phenomenon takes place only if the coupling constant $\al_g$ is greater than the critical value $\al_{c}$. In the case of graphene, although the coupling constant is about $2.2$ and it is assumed that the interaction is long-range, there is no experimental evidence for this phenomenon.
Theoretically, the computation of $\al_{c}$ has been the subject of extensive work (see e.g., [\onlinecite{Kotikov:2016yrn,Teber:2018jdh}] and references therein).
All articles seem to agree that $\al_c$ should be of order $1$.
However, at present there is still no consensus on the exact value of $\al_c$ and whether it is greater than $\al_g$. 
It is clear that the exact calculation of $\al_c$ is a challenging
theoretical problem that would yield
an accurate knowledge of the phase structure
of graphene.
It is hoped that, for example, a
precise knowledge of $\al_c$ would
allow fine-tuning of (artificial) graphene-like materials to open a breach in these systems in a controlled manner. This would be of extreme
practical importance, for example, for the development of graphene-based transistors~\cite{nevius2015}.

Understanding the above questions requires a {\it quantitative} theoretical understanding of the effect of electron-electron interactions. To achieve this goal, we will advocate the need for a complete understanding of the IR Lorentz invariant fixed point as a precondition for a correct understanding of the experimentally more relevant nonrelativistic limit [\onlinecite{Teber:2012de,Teber:2018jdh}].

Interestingly, the study of this fixed point brings us back to old problems, such as dynamic chiral symmetry breaking in 3D QED (QED$_3$), where the quantity of greatest interest is the critical number of flavors, $N_c$ (see e.g., [\onlinecite{Pisarski:1984dj}]
and also [\onlinecite{Kotikov:2016wrb,Gusynin:2016som,Teber:2018jdh}]).
QED$_3$ in $1/N$-expansion is a model closely related to the model describing graphene at a fixed point.
Indeed, such a correspondence arises because the photon propagators in both models have the same form $\sim 1/p$ (see Ref.~[\onlinecite{Teber:2019kkp}] and discussions
therein).

The critical coupling constant in the two extreme limits reads:
%
\be
0.833 < \al_c < 7.65, \qquad \al_c^* = 1.22\, ,
\label{models:alc-res}
\ee
where the range of values in the non-relativistic limit is taken from the literature on the subject, see [\onlinecite{Teber:2018jdh}],
and in the ultrarelativistic limit, the result has the next-to-leading (NLO) accuracy and is gauge-invariant.
One can also compare modern gauge-invariant $N_c$ NLO calculations for QED$_{4,3}$ \cite{Kotikov:2016yrn} and QED$_3$ \cite{Kotikov:2016wrb,Gusynin:2016som}:
%
\be
\text{QED}_{4,3}:~~N_c^*=3.04, \qquad \text{QED}_3:~~N_c=2.85\, .
\ee
%

{\bf Resume.} In this
short review, we considered the optical conductivity and the dynamic gap generation in graphene.\\

{\bf Acknowledgments.}
The author is grateful to Sofian Teber for help
in preparing the paper. He also thanks the Organizing Committee of the
International Conference “Modern problems of condensed matter theory” for
the invitation.


\end{fmffile}

\begin{thebibliography}{99}

 \bibitem{Semenoff:1984dq}
G.~W.~Semenoff,
\href{http://dx.doi.org/10.1103/PhysRevLett.53.2449}{Phys.\ Rev.\ Lett.\  {\bf 53} (1984) 2449}.

\bibitem{Gonzalez:1993uz}
J.~Gonz\'alez, F.~Guinea and M.~A.~H.~Vozmediano,
\href{http://dx.doi.org/10.1016/0550-3213(94)90410-3}{Nucl.\ Phys.\ B {\bf 424} (1994) 595}

\bibitem{Kotikov:2018wxe}
A.~V.~Kotikov and S.~Teber,
Phys. Part. Nucl. \textbf{50} (2019) no.1, 1-41

\bibitem{Gusynin:2006ym}
  V.~P.~Gusynin, S.~G.~Sharapov and J.~P.~Carbotte,
  \href{http://dx.doi.org/10.1103/PhysRevLett.96.256802}{Phys.\ Rev.\ Lett.\  {\bf 96} (2006) 256802};
  \href{http://stacks.iop.org/1367-2630/11/i=9/a=095013}{New Journal of Physics {\bf 11} (2009) 095013}

\bibitem{Teber:2014ita}
S.~Teber and A.~V.~Kotikov, 
\href{http://dx.doi.org/10.1209/0295-5075/107/57001}{Europhys.\ Lett.\  {\bf 107} (2014) 57001};
\href{http://dx.doi.org/10.1134/S004057791703014X}{Theor. Math. Phys.
  {\bf 190} (2017) 446};
\href{http://dx.doi.org/10.1007/JHEP07(2018)082}{JHEP {\bf 1807} (2018) 082}

\bibitem{Teber:2018jdh}
S.~Teber, 
\href{https://arxiv.org/abs/1810.08428}{Habilitation, Sorbonne Universit\'e (2017) [arXiv:1810.08428 [cond-mat.mes-hall]]}.



\bibitem{Fradkin86.PhysRevB.33.3263}
E.~Fradkin, 
\href{http://dx.doi.org/10.1103/PhysRevB.33.3263}{Phys.\ Rev.\ B {\bf 33} (1986) 3263};
P.\ A.~Lee, 
\href{http://dx.doi.org/10.1103/PhysRevLett.71.1887}{Phys.\ Rev.\ Lett.\ {\bf 71} (1993) 1887};
A.\ W.\ W.~Ludwig, M.\ P.\ A.~Fisher, R.~Shankar and G.~Grinstein, 
\href{http://link.aps.org/doi/10.1103/PhysRevB.50.7526}{Phys.\ Rev.\ B {\bf 50} (1994) 7526}.

\bibitem{Bogoliubov:1957gp}
  N.~N.~Bogoliubov and O.~S.~Parasiuk, 
\href{http://dx.doi.org/10.1007/BF02392399}{Acta Math.\  {\bf 97} (1957) 227};
  K.~Hepp, 
\href{http://dx.doi.org/10.1007/BF01773358}{Commun.\ Math.\ Phys.\  {\bf 2} (1966) 301};
W.~Zimmermann, 
\href{http://dx.doi.org/10.1007/BF01645676}{Commun.\ Math.\ Phys.\  {\bf 15} (1969) 208}.

\bibitem{Kotikov:2021hsy}
A.~V.~Kotikov,
Particles \textbf{4} (2021) no.3, 361-380.

\bibitem{Mishchenko2008}
E.\ G.~Mishchenko, 
\href{http://stacks.iop.org/0295-5075/83/i=1/a=17005}{Europhys.\ Lett.\ {\bf 83} (2008) 17005}


\bibitem{Kotikov:2016yrn}
  A.~V.~Kotikov and S.~Teber, 
  \href{http://dx.doi.org/10.1103/PhysRevD.94.114010}{Phys.\ Rev.\ D {\bf 94} (2016) no.11,  114010}

\bibitem{nevius2015}
  M.\ S.~Nevius, et al.
  \href{http://link.aps.org/doi/10.1103/PhysRevLett.115.136802}{Phys.\ Rev.\ Lett.\ {\bf 115} 136802 (2015)}

\bibitem{Teber:2012de}
S.~Teber, 
\href{http://dx.doi.org/10.1103/PhysRevD.86.025005}{Phys.\ Rev.\ D {\bf 86} (2012) 025005};
A.~V.~Kotikov and S.~Teber, 
\href{http://dx.doi.org/10.1103/PhysRevD.87.087701}{Phys.\ Rev.\ D {\bf 87} (2013) 087701};
\href{http://link.aps.org/10.1103/PhysRevD.89.065038}{Phys.\ Rev.\ D {\bf 89} (2014) 065038};
  S.~Teber and A.~V.~Kotikov,
  \href{http://dx.doi.org/10.1103/PhysRevD.97.074004}{Phys.\ Rev.\ D {\bf 97} (2018) no.7,  074004}

\bibitem{Pisarski:1984dj}
R.~D.~Pisarski, 
\href{http://dx.doi.org/10.1103/PhysRevD.29.2423}{Phys.\ Rev.\ D {\bf 29} (1984) 2423};
T.~Appelquist, D.~Nash and L.~C.~R.~Wijewardhana, 
\href{http://dx.doi.org/10.1103/PhysRevLett.60.2575}{Phys.\ Rev.\ Lett.\  {\bf 60} (1988) 2575};
D.~Nash, 
\href{http://dx.doi.org/10.1103/PhysRevLett.62.3024}{Phys.\ Rev.\ Lett.\  {\bf 62} (1989) 3024};
A.~V.~Kotikov, 
\href{http://www.jetpletters.ac.ru/ps/1193/article_18004.shtml}{JETP Lett. {\bf 58} (1993) 731};
\href{http://dx.doi.org/10.1134/S1063778812070058}{Phys.\ Atom.\ Nucl.\  {\bf 75} (2012) 890}

\bibitem{Kotikov:2016wrb}
A.~V.~Kotikov, V.~I.~Shilin and S.~Teber, 
\href{http://dx.doi.org/10.1103/PhysRevD.94.056009}{Phys.\ Rev.\ D {\bf 94} (2016) 056009};
  A.~V.~Kotikov and S.~Teber, 
  \href{http://dx.doi.org/10.1103/PhysRevD.94.114011}{Phys.\ Rev.\ D {\bf 94} (2016) no.11,  114011};
Particles \textbf{3} (2020) no.2, 345-354
  
\bibitem{Gusynin:2016som}
V.~P.~Gusynin and P.~K.~Pyatkovskiy, 
\href{http://dx.doi.org/10.1103/PhysRevD.94.125009}{Phys.\ Rev.\ D {\bf 94} (2016) no.12,  125009}

\bibitem{Teber:2019kkp}
S.~Teber and A.~V.~Kotikov,
Theor. Math. Phys. \textbf{200} (2019) no.2, 1222-1236
\end{thebibliography}
\end{document}